# Relating Energy Level Alignment and Amine-Linked Single Molecule Junction Conductance

*M. Dell'Angela[1,2], G. Kladnik[3], A. Cossaro[1], A. Verdini[1], M. Kamenetska[4], I. Tamblyn[5], S.Y. Quek[5],*

*J.B. Neaton[5], D. Cvetko[3], A. Morgante[1,2], L. Venkataraman[4]*

[1]CNR-IOM Laboratorio Nazionale TASC, Basovizza SS-14, km 163.5, I-34012 Trieste, Italy
[2]Department of Physics, University of Trieste, Trieste, Italy
[3]Dept. of Physics, University of Ljubljana, Ljubljana, Slovenia
[4]Dept. of Applied Physics and Applied Mathematics, Columbia University, New York, NY
[5]Molecular Foundry, Lawrence Berkeley National Laboratory, Berkeley, CA

morgante@tasc.infm.it; jbneaton@lbl.gov; lv2117@columbia.edu

**Abstract**: Using photoemission spectroscopy, we determine the relationship between electronic energy level alignment at a metal-molecule interface and single-molecule junction transport data. We measure the position of the highest occupied molecular orbital (HOMO) relative to the Au metal Fermi level for three 1,4-benzenediamine derivatives on Au(111) and Au(110) with ultraviolet and resonant x-ray photoemission spectroscopy. We compare these results to scanning tunnelling microscope based break-junction measurements of single molecule conductance and to first-principles calculations. We find that the energy difference between the HOMO and Fermi level for the three molecules adsorbed on Au(111) correlate well with changes in conductance, and agree well with quasiparticle energies computed from first-principles calculations incorporating self-energy corrections. On the Au(110) which present Au atoms with lower-coordination, critical in break-junction conductance measurements, we see that the HOMO level shifts further from the Fermi level. These results provide the first direct comparison of spectroscopic energy level alignment measurements with single molecule junction transport data.

Understanding the electronic structure of individual molecules bonded between metal electrodes is critical to progress in molecular electronics[1]. Typically, the electronic structure of molecular junctions is inferred from low-bias conductance measurements[2-6]. However, transport measurements are at best indirect probes of the alignment of the molecular orbitals to the metal Fermi level. While the ionization potential and electron affinity of molecules can be routinely characterized and calculated in the gas-phase, additional physical effects, such as charge transfer and rearrangement, hybridization, and electrode polarization are expected to drastically alter these electronic removal and addition energies in molecular junctions. In fact, it is well known that at the interfaces of organic semiconductors and metals, the vacuum level alignment rule breaks down[7], and thus gas-phase electronic structure cannot be used to quantitatively determine level alignments in molecular junctions. Although trends in zero-bias conductance as a function of substituents[8, 9], oxidation potential[10] or thermo-electric voltage measurements[11, 12] have been used to infer the nature of transport (hole or electron) through molecular junctions, spectroscopic measurements are required to probe level alignments directly[13, 14]. Quantitative measurement of junction level alignment is currently beyond the resolution of standard photoemission spectroscopies. However, photoemission from molecules adsorbed to metal contact surfaces, a related system, is accessible. As we shall show, combined with conductance measurements and a first-principles theory capable of accounting for the physical effects outlined above, an explicit connection between single-molecule junction conductance and level alignment can be obtained.

Here, we use photoemission spectroscopy to measure quantitatively the position of the highest occupied molecular orbital (HOMO) energy of three 1,4-benzenediamine derivatives relative to the Fermi energy on two Au surfaces (Au(111) and Au(110)). We find excellent quantitative agreement between these results and those from first-principles density functional theory (DFT) calculations including self-energy corrections. Comparing our results on Au(110) with Au(111), we find that HOMO levels shift further away from $E_F$ on the Au(110) surface, while the adsorption energy of the molecules on the metal surface increases, as expected due to the decrease in coordination of the surface atoms on Au(110). Finally, comparing these molecule-on-surface measurements and calculations to single molecule conductance data[8] from scanning tunnelling microscope (STM) based break-junction measurements we find that the shifts in the molecular levels correlate well with changes in conductance values.

We focus here on three amine-terminated molecules: tetramethyl-1,4-diaminobenzene (**TMBDA**), which has four electron donating methyl substituents; 1,4-diaminobenzene (**BDA**); and tetrafluoro-1,4-diaminobenzene (**TFBDA**), which has four electron withdrawing fluorine substituents (all are from commercial sources). We use ultraviolet photoemission spectroscopy (UPS) and resonant x-ray photoemission spectroscopy (resonant XPS) to study the electronic properties of monolayer coverage on Au(111) and Au(110), and near edge X-ray absorption fine structure (NEXAFS) to determine the molecule orientation on these surfaces. The measurements are performed at the ALOISA/HASPES beamline (Elettra Synchrotron, Trieste)[15-17].

Monolayers of these three molecules are prepared on Au(111) and first characterized using Helium Atom Scattering (HAS) measurements. Thick films of molecules on Au(111) are first grown at 270K. The sample temperature is then ramped at a constant heating rate of about 6 K/min, during which the Helium specular reflectivity is measured (Figure 1). The HAS signal is close to zero when the Au(111) surface is covered with a thick film and increases with increasing temperature as molecules desorb until it reaches a maximum when the surface is clean. When an ordered monolayer is formed, the HAS signal is typically higher than that on a multilayer film. This may be observed for a molecule dependent range of temperatures (see SI Figure S2), and as the temperature is increased further, the HAS signal starts to increase steeply, as molecules of the monolayer film desorb exposing the Au substrate beneath. The temperature at which the three molecules desorb from the Au(111) surface[18] is observed to be 297±15 K, 327±15 K and 415±15 K for **TFBDA, BDA** and **TMBDA** respectively, and monolayer coverages of the three molecules on Au(111) are obtained from a multilayer film by heating the samples to 260K, 270K and 300K respectively, where these temperature are chosen within the region with approximately constant HAS signal. Since these molecules bind to gold through an N-Au donor-acceptor bond[6, 19], we expect electron withdrawing (donating) substituents on the benzene ring to weaken (strengthen) this bond[8]. Indeed, this agrees with estimated the molecular adsorption energies of 0.9 eV, 1.0 eV and 1.2 eV for **TFBDA, BDA** and **TMBDA** respectively calculated using Redhead formula[20], assuming first order desorption kinetics[21].

NEXAFS measurements are performed at the C K-edge in order to determine the orientation of the molecule on the surface. In Figure 2, we show the partial electron yield (PEY) of the C K-edge NEXAFS for each molecular monolayer, acquired with the light polarization parallel with respect to the surface (s-pol) and perpendicular with respect to the surface (p-pol). The first two sharp peaks are assigned to the C1s ⊠ $\pi^*$ excitation, which are localized on the benzene ring[22, 23]. The intensity of these two peaks depends on the orientation of the benzene ring on the surface, and the angle of the light polarization vector with respect to the surface. It is maximal when the polarization is orthogonal to the plane of the ring. Thus, from the change in intensity of these two peaks as a function of the light polarization angle[24], we determine the angle θ of the benzene rings with respect to the surface to be 27°±10°, 24°±10° and 12°±10° for **TMBDA**, **BDA** and **TFBDA** respectively.

UPS spectra (HeI, hν=21.2eV) for monolayers of the three molecules on Au(111) acquired at normal emission are shown in Figure 3a (dots). The HOMO peaks are determined from best-fits to the raw data which are also shown (see SI for details). Compared with **BDA**, we find that the HOMO peak is shifted closer to the Au Fermi level by ~0.4eV for **TMBDA**, but shifted away from the Au Fermi level by ~0.1eV (relative to **BDA**) for **TFBDA** (see Table 1). To elucidate features of the valence band of the TMBDA, BDA and TFBDA monolayers films, we have performed resonant XPS measurements, which allows us to distinguish molecular features from the valence band of the Au substrate, which dominates the UPS spectra[25-27]. In resonant XPS, the energy of the incident photon is tuned to the energy corresponding to the peaks in the NEXAFS spectra, and thus an electron from the core (from the C1s orbital) is first promoted to an unoccupied molecular state (LUMO, LUMO+1 ...). When the decay of this electron results in an Auger-like electron emission from the molecular occupied orbitals, a resonant enhancement is achieved. Thus in resonant XPS, an additional electron emission channel opens from the occupied orbitals due to a resonant process involving a core-hole excitation[26, 28, 29].

Figure 3b shows a series of valence band photoemission spectra taken with photon energies of ν=286.6 eV, 286.6 eV and 287 eV for **TMBDA**, **BDA** and **TFBDA** respectively. These energies correspond to the $C_{1,4}$ 1s → $\pi_{C-N}^*$ transitions in the C K-edge NEXAFS, i.e. from $C_{1,4}$ 1s to the LUMO+1. Only the resonant contribution of the measured occupied orbitals spectra for the three molecules is shown in Figure 3b. The non-resonant part, measured at a photon energy of (hν=284 eV) has been subtracted from the raw. Distinct valence band peaks are identified in these spectra for all three molecules, both on the Au(111) and Au(110) surfaces. The chosen photon energies show clearly the HOMO peak position for different systems.

Both UPS and resonant XPS measurements find that **TMBDA** has the HOMO closest from $E_F$, while **TFBDA** has the HOMO furthest from $E_F$. We note here a slight difference of the HOMO positions measured with UPS and resonant XPS. This is primarily due to the occupation of different vibrational sublevels of the electronic final state in the resonant XPS[30]. Notwithstanding these small differences the sequence and relative energies of the HOMO peaks are the same for both methods, giving confidence in our results reported in Table 1. Comparing the HOMO positions relative to $E_F$ across these three molecules with single-molecule break-junctions conductance data[8], where the molecules are bonded to Au at both amine end-groups (see Table 1) we see that a deeper HOMO correlates with a smaller conductance. Interestingly, for both the conductance and photoemission data, the effect of four electron-withdrawing F groups on the HOMO position is not the same as that of four electron-donating $CH_3$ groups, although they have Hammett parameters of similar magnitude[31].

On Au(110), resonant XPS measurements place the HOMO energies deeper relative to Fermi, by about ~0.3eV when compared $E_F$-$E_{HOMO}$ difference on the Au(111). This can be attributed partly to the difference in the nature of the adsorption site between the two surface: the (110) surface Au atoms have a lower coordination when compared with the (111) surface Au atoms. Indeed, we find, from HAS measurements, that these molecules are more strongly bound on Au(110) (see Supporting Information). However, the ~ 0.3eV shift in the HOMO position on Au(110) could also result from a difference in the screening between the two surfaces[32].

In order to understand further the correlation between spectroscopic and conductance measurements, we perform first-principles DFT calculations, within the generalized gradient approximation (GGA-PBE)[33], as implemented in VASP[34]. We use a periodic supercell with ~10 Å of vacuum and a six-layer gold slab, with each layer consisting of 16 atoms (4x4 unit cell) to model a (111) surface. The bottom three Au layers are frozen in their bulk positions and all other atoms are allowed to relax until forces are converged to within 50 meV/Å. The planewave cutoff is 435 eV, and the electronic charge density is determined self-consistently in the presence of an external electrostatic potential which removes artifacts introduced when using periodic boundary conditions for adsorbates on one side of the metal slab.[35] The molecular orbital energy level alignments are calculated within a self-energy corrected DFT-GGA framework, implemented in SIESTA[36], as described below[19]. We use k-grids of 2x2x1 and 5x5x1 to calculate the charge density and electronic density of states respectively. The DFT-PBE HOMO energies computed in this manner are within 0.1 eV of those obtained with VASP.

We optimize geometries for all three molecules on the Au(111) surface and find the Au-N-C angle, $\theta$, to be 23°, 27°, and 54° for **TFBDA**, **BDA**, and **TMBDA**, with GGA-PBE adsorption energies of 0.26 eV, 0.36 eV, and 0.44 eV, respectively[37, 38] (see SI for details). In all three cases, the molecule binds to the planar Au(111) surface via the atop site, with Au-N bond lengths of 2.8, 2.6, and 3.1 Å. Our optimized angles differ from those measured in experiment, probably because van der Waals interactions, which are of the scale of our computed adsorption energies, are not described within GGA-PBE. In what follows, we compute level alignment for molecules relaxed with $\theta$ fixed to the value reported in experiment as shown in Figure 4.

For a given adsorbate geometry, energy level alignment is determined by evaluating the site-projected partial density of states (PDOS) for atoms on the molecule. Since DFT underestimates electron removal/addition (quasiparticle) energy gaps between the highest occupied (HOMO) and lowest unoccupied (LUMO) orbital energies, molecular resonances are predicted far too close to the metal Fermi level, $E_F$, resulting in poor agreement with spectroscopy experiments. To correct for DFT level alignment and compare quantitatively with photoemission data, we employ a physically motivated electron self-energy correction to the molecular orbital energies at the surface (DFT+Σ)[19, 39] which consists of two parts: a "bare or molecular" term, correcting for the DFT HOMO-LUMO gap of the gas-phase molecule, computed from total energy differences; and second, an "image-charge" term which accounts for the effect of electrode polarization on the energy of the added electron or hole[40]. (See SI for details.) This interface-dependent self-energy has no adjustable parameters and is expected to give an accurate description of quasiparticle level alignment for weakly coupled molecule–metal substrate states for which frontier orbital character is not significantly altered from their gas-phase counterparts through coupling to the surface.

Using the DFT+Σ approach, the HOMO energy levels, relative to $E_F$, are calculated to be -1.2 eV, -1.6 eV, and -1.8 eV for **TMBDA**, **BDA** and **TFBDA**, respectively, for the experimentally-determined geometries. These results are in excellent quantitative agreement with our photoemission data summarized in Table 1. In the absence of self-energy corrections at this coverage, the DFT HOMO energy levels are computed to be -0.1, -0.4, -0.5 eV for **TMBDA**, **BDA** and **TFBDA** respectively, where although the trends are predicted correctly, the agreement with experimental data is extremely poor. Furthermore, this disagreement between the experiment and DFT HOMO levels cannot be accounted for with changes to the angle θ within the ±10° experimental error[41], or with changes to the molecular coverage, as we have explored different molecular coverages using three different unit cells (3x3, 4x4, 5x5). We find that these changes in coverage lead to variations in HOMO of about ±0.2 eV, and as coverage increases, the surface (or bond) dipole decreases and the HOMO energy moves further away from $E_F$.

Interestingly, our DFT+Σ calculations predict a much deeper HOMO level of about -3.4 eV in the Au-**BDA**-Au junction geometry (where **BDA** is bound to adatoms at both amine groups)[19, 39] compared with -1.6 eV for the **BDA** on the atop site of the Au(111) surface. This large difference has three main physical origins. First, binding to undercoordinated Au adatoms results in larger charge transfer from **BDA** to Au: at the DFT level, when **BDA** binds to an adatom rather than the atop site on Au(111), the HOMO deepens from -0.4 eV to -0.6 eV. Second, binding to adatoms on Au(111) in a junction geometry further deepens the DFT HOMO to -1.1 eV, reflecting more charge transfer due to binding at both amines. Third, the image-charge self-energy corrections, which depend on the position and orientation of **BDA** relative to the image plane, can be smaller in the junction, resulting in a deeper HOMO. Thus transport measurements through junctions where the molecule is bonded only on one surface can yield results that are significantly different form measurements in which molecules are bonded to two electrodes as long as the transport mechanism is the same in both cases[42].

In conclusion, we determine the HOMO energy levels for three different 1,4-diaminobenzene derivatives on Au(111) using ultraviolet and resonant x-ray photoemission spectroscopy, and find excellent quantitative agreement between these measurements and self-energy corrected DFT calculations. We find that the trends in the measured level alignment correlate well with single molecule conductance data, indicating that the trend for the HOMO energy levels in a junction geometry is consistent with that on Au(111). Furthermore, we note that the measured molecular adsorption energies increase from **TFBDA** to **BDA** to **TMBDA**. These results thus provide the first direct comparison between energy level alignment and single molecule transport measurements.

**Acknowledgements:** This work was supported in part by the NSEC program of the NSF (Grant No. CHE-0641523), and a NSF Career grant (No. CHE-07-44185). Portions of this work were performed at the Molecular Foundry, Lawrence Berkeley National Laboratory, and were supported by the Office of Science, Office of Basic Energy Sciences, of the US Department of Energy. AM gratefully acknowledges the NSEC at Columbia University and the Italian Academy at Columbia University for the warm hospitality and financial support during his visit. Current address for MD is Institut für Experimentalphysik, Universität Hamburg, 22761 Hamburg.

**Supporting Information Available:** Experimental methods and theoretical calculations details.

Table 1:

| Molecule | Measured Molecule Adsorption Energy [eV] | Calculated Molecule Adsorption Energy [eV] | HOMO position relative to $E_F$ [eV] | | | | Experimental Conductance $10^{-3} G_0$ |
|---|---|---|---|---|---|---|---|
| | | | Au(111) | | | Au(110) | |
| | | | UPS | XPS | Theory | XPS | |
| TMBDA | 1.2 | 0.44 | -1.0 ± 0.1 | -1.3 ± 0.2 | -1.2 | -1.6 ± 0.2 | 8.2 ± 0.2 |
| BDA | 1 | 0.36 | -1.4 ± 0.1 | -1.6 ± 0.2 | -1.6 | -1.8 ± 0.2 | 6.4 ± 0.2 |
| TFBDA | 0.9 | 0.26 | -1.5 ± 0.1 | -1.65 ± 0.2 | -1.8 | _ | 5.5 ± 0.2 |

**Table 1:** Measured molecule adsorption energy on Au(111) determined from temperature dependent HAS measurement, calculated molecule adsorption energy on Au(111) using optimized molecule geometries, HOMO energy level relative to $E_F$, as determined experimentally from UPS and resonant XPS on Au(111) and Au(110), and determined theoretically. Conductance values as determined in single molecule break junction experiments[8].

Figure 1:

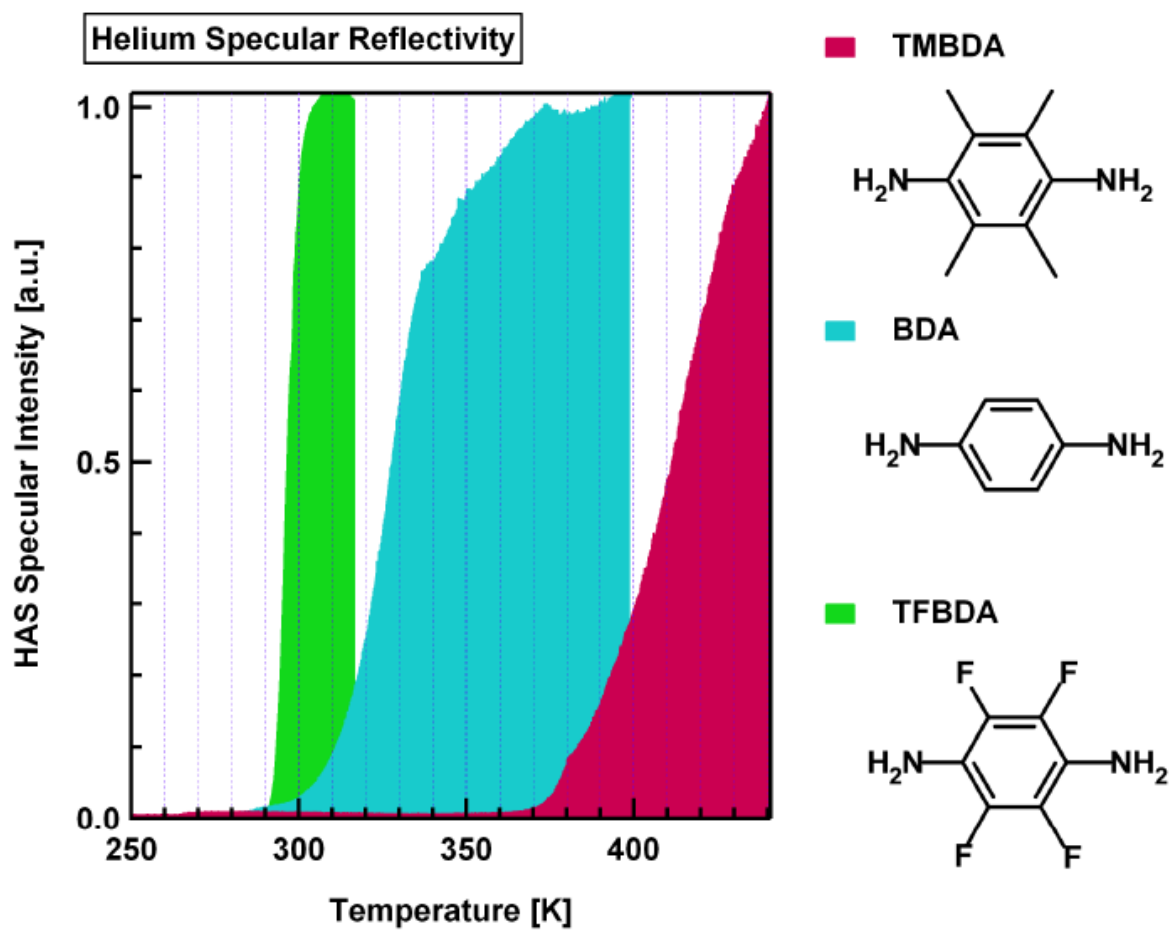

**Figure 1**. Helium specular reflectivity curves measured during thermal desorption of the thick films of tetramethyl-1,4-benzenediamine (TM**BDA**), 1,4-benzenediamine (**BDA**) and tetrafluoro-1,4-benzenediamine (TF**BDA**) on Au(111). The reported desorption temperature is defined as the point when the HAS signal has reached half its maximal value.

Figure 2:

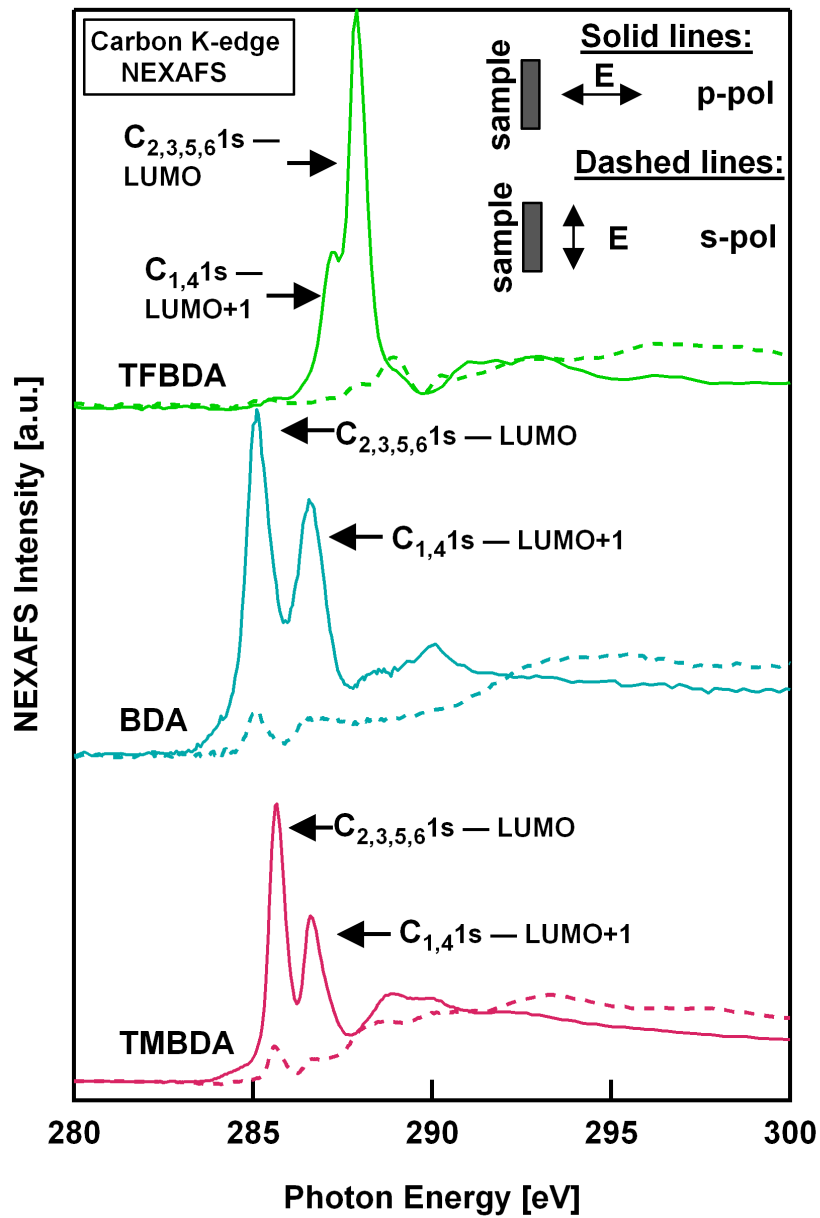

**Figure 2:** Partial electron yield C K-edge NEXAFS for the monolayers of the three molecules on Au(111) acquired with the light polarization parallel/perpendicular with respect to the surface (dotted /solid lines). The benzene rings tilt angles obtained from NEXAFS linear dichroisms are 27°±10°, 24°±10° and 12°±10° for **TMBDA**, **BDA** and **TFBDA,** respectively.

Figure 3:

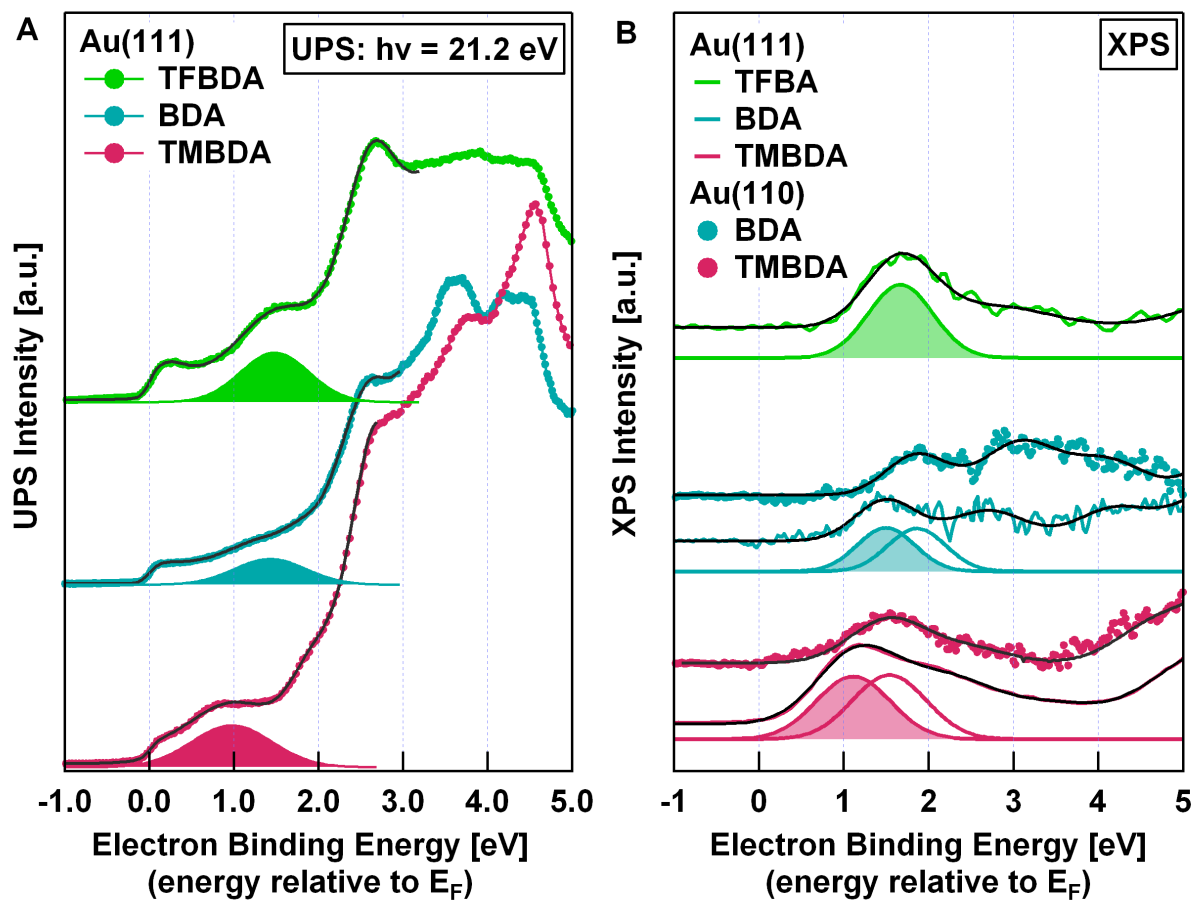

**Figure 3:** UPS (hν=21.2eV) and resonant XPS (hν=286.6eV for **BDA** and TM**BDA** and hν=287eV for TF**BDA**) valence band measurements of thin molecular films on Au(111) and Au(110). The solid black lines are the best fits and the HOMO peaks resulting from the best fitting procedure are reported at the bottom of each trace (see SI for details).

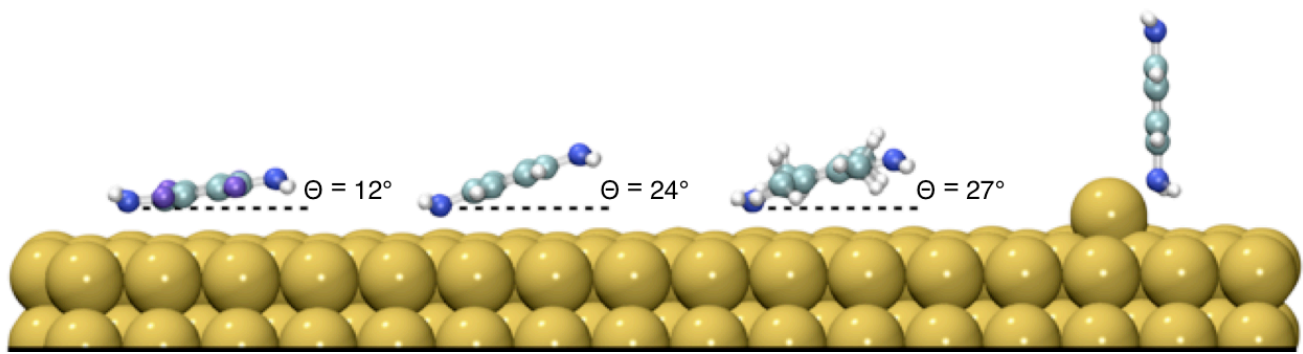

**Figure 4**: Left to right: Atomic geometries for TM**BDA**, **BDA**, TF**BDA** on Au(111) for molecular angles determined from NEXAFS analysis and **BDA** bound to an adatom.